\documentclass{article}
\usepackage{spconf,amsmath,graphicx}
\usepackage{epstopdf}
\usepackage{url}
\usepackage{amsfonts}
\usepackage{amsmath}
\usepackage{bm}
\usepackage{algorithm}
\usepackage{algorithmic}
\usepackage{array}
\usepackage{amssymb}
\usepackage{booktabs}
\usepackage{makecell}
\usepackage{multirow}
\usepackage{subfigure}

\title{Boosting Noise Robustness of Acoustic Model via Deep Adversarial Training}
\small \name{Bin Liu$^{1,2}$ \qquad Shuai Nie$^{1,2,*}$ \qquad Yaping Zhang$^{1,2}$ \qquad Dengfeng Ke$^{1,3}$ \qquad Shan Liang$^1$ \qquad Wenju Liu$^1$ \thanks{* denotes equal contribution to this work.}}
 \address{$^1$ National Laboratory of Patten Recognition, Institute of Automation, Chinese Academy of Sciences, China\\
  $^2$ School of Artificial Intelligence, University of Chinese Academy of Sciences, China\\
  $^3$ School of Information Science and Technology, Beijing Forestry University, China\\
  \small\texttt{\{bin.liu2015,shuai.nie,yaping.zhang,dengfeng.ke,sliang,lwj\}@nlpr.ia.ac.cn}}

\begin{document}
%
\maketitle
\begin{abstract} \small
In realistic environments, speech is usually interfered by various noise and reverberation, which dramatically degrades the performance of automatic speech recognition (ASR) systems. To alleviate this issue, the commonest way is to use a well-designed speech enhancement approach as the front-end of ASR. However, more complex pipelines, more computations and even higher hardware costs (microphone array) are additionally consumed for this kind of methods. In addition, speech enhancement would result in speech distortions and mismatches to training. In this paper, we propose an adversarial training method to directly boost noise robustness of acoustic model. Specifically, a jointly compositional scheme of generative adversarial net (GAN) and neural network-based acoustic model (AM) is used in the training phase. GAN is used to generate clean feature representations from noisy features by the guidance of a discriminator that tries to distinguish between the true clean signals and generated signals. The joint optimization of generator, discriminator and AM concentrates the strengths of both GAN and AM for speech recognition. Systematic experiments on CHiME-4 show that the proposed method significantly improves the noise robustness of AM and achieves the average relative error rate reduction of 23.38\% and 11.54\% on the development and test set, respectively.

\end{abstract}
\begin{keywords} \small
robust speech recognition, deep adversarial training, acoustic model, generative adversarial net
\end{keywords}
\section{Introduction} \small
\label{sec:intro}

In realistic environments, speech is usually interfered by various noise and reverberation, which dramatically degrades the performance of automatic speech recognition (ASR) systems. Many neural network-based acoustic model training schemes have been proposed to boost the noise robustness of ASR, such as noise-aware training \cite{seltzer2013investigation}, noise adaptive training (NAT) \cite{narayanan2014joint} and DNN adaptation \cite{liao2013speaker}. Various elaborate acoustic model architectures have also been proposed in order to  promote the modeling capability. However, the complicated networks have high computational complexities and the degraded speech might lack its crucial discriminative characteristics, which makes it hard to significantly improve the ASR performance.

Another mainstream approach to boost noise robustness is adding a well-designed speech enhancement part during the front-end of ASR. The processing methods include traditional statistical methods like Wiener filter \cite{Lim1978All} and STSA-MMSE \cite{Loizou2013Speech}, a denoising autoencoder (DAE) \cite{vincent2008extracting} and DNN-based ideal binary masking \cite{li2013improving}. However, more complex pipelines,  more computations and even higher hardware costs (microphone array) are additionally consumed for this kind of methods. In addition, these enhancement methods use a particular form of loss functions such as mean squared error, which tends to generate over-smoothed spectra that lack the fine structures that are near to those of the true speech. The speech distortions and mismatches to training sometimes degrade the ASR performance. Moreover, it fails to optimize towards the final objective since the speech enhancement part is usually distinct from the recognition part, which leads to a suboptimal solution \cite{Seltzer2008Bridging}.

Generative adversarial nets (GANs) introduced by Goodfellow et al. \cite{Goodfellow2014Generative} have attracted a lot of attention, which aims to generate samples following the underlying data distribution without the need for the explicit form of distribution. In addition, to obtain an optimal performance, a joint training of speech enhancement and acoustic model is proposed for robust speech recognition \cite{gao2015joint}.

In this paper, we propose an adversarial training method to directly boost noise robustness of acoustic model. Specifically, a jointly compositional scheme of generative adversarial net (GAN) and neural network-based acoustic model (AM) is used in the training phase. GAN is used to generate clean feature representations from noisy features by the guidance of a discriminator that tries to distinguish between the true clean signals and generated signals. Without the complex front-end approaches and the need for one-to-one correspondence between the true clean and generated samples, our method greatly simplifies the training process. The use of adversarial training circumvents the limitation of hand-engineering loss functions and captures the underlying structural characteristics from the noisy signals, which could boost noise robustness of acoustic model.

This paper is organized as follows. The related work is discussed in Section \ref{sec:prior}. The basic and conditional generative adversarial nets are presented in Section \ref{sec:GANs} . The proposed deep adversarial training is described in Section \ref{sec:Speech GAN}. The experimental results and conclusions are given in Section \ref{sec:Experiments} and Section \ref{sec:conclusions}, respectively.

\section{RELATION TO PRIOR WORK}
\label{sec:prior}

Generative adversarial nets (GANs) have attracted a lot of attention recently because of their successful applications in the computer vision and image processing tasks \cite{berthelot2017began}. GANs have also been applied to speech conversion \cite{hsu2017voice}, synthesis \cite{Bollepalli2017Generative} and enhancement \cite{Pascual2017SEGAN} tasks. The use of GANs as classifiers has already been investigated for image classification tasks \cite{springenberg2015unsupervised,Odena2016Semi}, which extends GANs to learn  discriminative classifiers by forcing the discriminator to output class labels. Shen et al.\cite{shen2017conditional} used conditional generative adversarial nets as a classifier for the spoken language identification task. However, it's hard to achieve the equilibrium for a single discriminator network which plays two competing roles of fake samples identifier and labels predictor. Recently, Li et al. \cite{li2017triple} proposed triple-GAN to introduce three components for both classification and class-conditional generation. To the best of our knowledge, using GANs for robust speech recognition has not yet been studied, so our method is the first approach to use the adversarial training framework for robust speech recognition.

\section{Generative Adversarial Networks}
\label{sec:GANs}

GANs are generative models introduced by Goodfellow et al \cite{Goodfellow2014Generative}, which consist of a generator (G) and a discriminator (D). The generator G produces samples from the data distribution $P(\mathbf{x})$ by transforming noise variables $\mathbf{z}$ into fake samples $G(\mathbf{z})$. The discriminator D is a classifier that aims to recognize whether the sample is from G or training data. G is trained to produce outputs that cannot be distinguished from ``real" data by an adversarially trained D, which is trained to do as well as possible in detecting the generator's ``fakes". More formally, this adversarial learning process is formulated as a two-player minimax game with the objective
\begin{equation} \small
\label{gan}
\begin{aligned}
\min \limits_{G}\max \limits_{D} V (G,D ) &= \mathbb{E}_{ \mathbf{x} \sim p_{\textrm{data}}(\mathbf{x})}\left [ \log D (\mathbf{x}) \right ] \\
& + \mathbb{E}_{\mathbf{z} \sim p_{\mathbf{z}}(\mathbf{z})} \left [ \log \left ( 1 - D \left ( G ( \mathbf{z}) \right ) \right ) \right ].  \\
\end{aligned}
\end{equation}

Regular GANs suffer from the vanishing gradients problem because of the sigmoid cross-entropy loss function adopted for the discriminator. The least-squares GANs (LSGANs) approach \cite{Mao2016Least} substitutes the cross-entropy loss by the least-squares function, which can generate higher quality samples and perform more stable during the learning process. The objective functions for LSGANs can be defined as follows:
\begin{equation} \small
\label{lsgan}
\begin{aligned}
\min \limits_{D} V_{\textrm{LSGAN}} (D) &= \frac{1}{2} \mathbb{E}_{ \mathbf{x},\mathbf{x}_c \sim p_{\textrm{data}}(\mathbf{x},\mathbf{x}_c)}\left [ (D (\mathbf{x},\mathbf{x}_\textrm{c})-1)^{2} \right] \\
& + \frac{1}{2} \mathbb{E}_{\mathbf{z} \sim p_{\mathbf{z}}(\mathbf{z}), \mathbf{x}_c \sim p_{\textrm{data}}(\mathbf{x}_c)} \left [ ( D ( G ( \mathbf{z},\mathbf{x}_\textrm{c}),\mathbf{x}_\textrm{c} ) )^{2} \right ]  \\
\min \limits_{G} V_{\textrm{LSGAN}} (G) &= \frac{1}{2} \mathbb{E}_{\mathbf{z} \sim p_{\mathbf{z}}(\mathbf{z}), \mathbf{x}_c \sim p_{\textrm{data}}(\mathbf{x}_c)} \left [ ( D ( G ( \mathbf{z},\mathbf{x}_\textrm{c}),\mathbf{x}_\textrm{c} ) -1)^{2} \right].
\end{aligned}
\end{equation}

\section{Deep Adversarial Training for Robust Speech Recognition}
\label{sec:Speech GAN} 

\begin{figure}[htb]
  \centering
  \includegraphics[width=0.8\linewidth ,trim=5 14 13 10,clip]{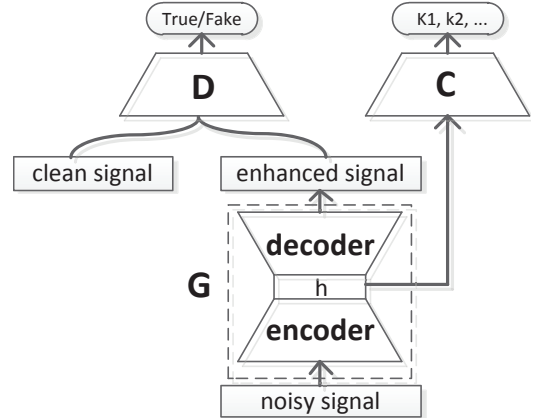}\\
  \caption{\small The structure of proposed adversarial training method: G is used to do speech enhancement by transforming the noisy speech signals into the enhanced version, D aims to distinguish between the clean and the enhanced samples, and C is to recognize the sensones.}
  \label{adn}
\end{figure}

For robust speech recognition, the main task of the acoustic model is to classify thousands of fine-grained senones given an input feature. We propose to train an acoustic model by the deep adversarial training method. The model consists of a generator(G), a discriminator (D) and a classifier(C), which is shown in the Fig.~\ref{adn}. In our case, the generator G performs the speech enhancement. It transforms the noisy speech signals into the enhanced version. The discriminator D aims to distinguish between the enhanced signals and clean ones. The classifier C classifies senones by features derivated from G. The encoding parts of G and the classifier C can be regarded as an organic whole of acoustic model.

The generator G is an encoder-decoder, adapted from \cite{Pascual2017SEGAN}. In the encoding stage, the input is passed and compressed through a series of strided convolutional layers followed by the leaky rectified linear units (LeakyReLUs) \cite{maas2013rectifier}. We choose the strided convolutions as they were shown to be more stable than other pooling approaches for GANs training \cite{Radford2015Unsupervised}. Downsample is done until we get a condensed bottleneck representation, called the hidden vector $\mathbf{h}$. The encoding process is reversed in the decoding stage by means of transposed convolutions, followed again by LeakyReLUs.

\begin{figure}[tb]
  \centering
  \includegraphics[width=0.6\linewidth ,trim=4 5 17 10,clip]{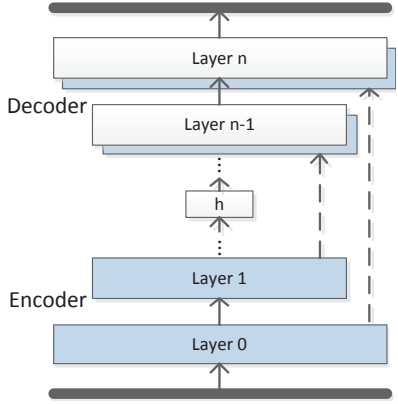}\\
  \caption{ \small ``U-Net" architecture. It is an encoder-decoder with skip connections between mirrored layers in the encoder and decoder stacks. The dashed arrows denote skip connections.}
  \label{unet}
\end{figure}

For speech enhancement, there is a great deal of low-level information shared between the input and output, and it would be desirable to shuttle the information directly across the model. For example, the noisy and clean speech share the same underlying structure. Many low-level details could be lost to reconstruct the speech signal if we force all information to flow pass through the bottleneck layer. Therefore, we add skip connections following the general shape of a ``U-Net" \cite{Ronneberger2015U}(Fig.~\ref{unet}). Specifically, each skip connection simply concatenates all channels at layer $i$ with those at layer $n-i$, where $n$ is the total number of layers. In addition, they are easier to optimize, as the gradients can flow deeper through the whole model \cite{He2016Deep}.

The discriminator D is built to model the high frequencies of the data and it gets the true clean samples from the dataset and generated samples from the generator as input. Differing from other GAN works, it does not use $\mathbf{z}$. Isola et al.\cite{Isola2016Image} report that adding a Gaussian noise as an input to G is not effective. We mention that there is no need for one-to-one correspondence between the true clean and generated samples.

The classifier C classifies senones and its input is the hidden vector $\mathbf{h}$ derivated from G. The bottleneck feature $\mathbf{h}$ is the compression of the noisy signal and captures the crucial structural characteristics with the guidance of D. Moreover, the classification progress is helpful to maintain more discriminative information for speech enhancement.

The G, D and C networks are jointly optimized by the adversarial training algorithm. Mathematically, given the noisy speech signal $\tilde{\mathbf{x}}$ and clean signal $\mathbf{x}$, G transforms it into enhanced version $\hat{\mathbf{x}}=G(\tilde{\mathbf{x}}; \theta_{g} )$. D tries to classify a signal as the clean or enhanced. With the LSGANs approach(Eq.~\ref{lsgan}), the formulation is
\begin{equation}
\label{vd}
\begin{aligned}
\min \limits_{D} V (D) &= \frac{1}{2} \mathbb{E}_{ \mathbf{x} \sim p_{\textrm{data}}(\mathbf{x})}\left [ (D (\mathbf{x})-1)^{2} \right] \\
& + \frac{1}{2} \mathbb{E}_{\tilde{\mathbf{x}} \sim p_{\textrm{data}}(\tilde{\mathbf{x}})} \left [ ( D ( G ( \tilde{\mathbf{x}}) ) )^{2} \right ].
\end{aligned}
\end{equation}
The input of the classifier C is the hidden vector $\mathbf{h}$, which computes the posterior probabilities over HMM states. K classes labels $\{k_1, k_2, ...k_K \}$ correspond to the state-level forced alignment targets. C is optimized by minimizing the category loss
\begin{equation}
\label{vc}
 \min \limits_{C} V(C) = \mathbb{E}_{\mathbf{h} \sim p(\mathbf{h})} \left [ - \log (C(k | \mathbf{h} ))\right].
\end{equation}
The generator G is trained to produce outputs that cannot be distinguished from ``real" samples by the discriminator D. In this way, D is in charge of transmitting information to G of what is real and what is fake, such that G can correct its output towards the realistic distribution. The adversarial G loss, with the LSGANs approach(Eq.~\ref{lsgan}), becomes
\begin{equation}
\label{vg}
\min \limits_{G} V_{\textrm{GAN}} (G) = \frac{1}{2} \mathbb{E}_{\tilde{\mathbf{x}} \sim p_{\textrm{data}}(\tilde{\mathbf{x}})} \left [ ( D ( G ( \tilde{\mathbf{x}}) ) -1)^{2} \right].
\end{equation}
In addition, the classifier C can also guide the generator to learn more discriminative information in the reconstruction of the speech signals. Therefore, the final G loss is
\begin{equation}
\label{vgall}
\min \limits_{G} V(G) = \alpha V_{\textrm{GAN}} (G) +  V(C).
\end{equation}
The magnitude of the adversarial loss is controlled by a hyper-parameter $\alpha$. When $\alpha = 0$, the model is equivalent to a traditional deep neural network model.

As shown in algorithm \ref{algorithm:adversarial-training}, we alternatively train the parameters of D, G and C to fine-tune the model. Three components are implemented with neural networks and the parameters are updated by stochastic gradient descent.

\begin{algorithm}[htb] \small
        \caption{\small The Deep Adversarial Training Algorithm }
        \label{algorithm:adversarial-training}
        \begin{algorithmic}[1]
        \REQUIRE Dataset $(\mathbf{x},\tilde{\mathbf{x}}) \in (\mathbf{X},\tilde{\mathbf{X}})$, $\mathbf{k}_i \in \mathbf{K}$, randomly initialized a generator $G$, a classifier $C$, a discriminator $D$ parameterized by $\theta_g, \theta_c,\theta_d$, learning rate $\mu$
        \ENSURE the optimized G, C and D parameterized by $\hat{\theta}_g, \hat{\theta}_c,\hat{\theta}_d$
        \STATE $//$ the adversarial training
        \REPEAT
        \FOR{number of training epochs}
            \FOR{ number of mini-batches}
                \STATE $ //$ for discriminator
                \STATE $  \theta_d \leftarrow \theta_d-\mu \frac{\partial V(D)}{\partial \theta_d} $
                \STATE $ //$ for generator
                \STATE $  \theta_g \leftarrow \theta_g-\mu \frac{\partial V(G)}{\partial \theta_g} $
                \STATE $ //$ for classifier
                \STATE $  \theta_c \leftarrow \theta_c-\mu \frac{\partial  V(C)}{\partial \theta_c} $
           \ENDFOR
        \ENDFOR
        \UNTIL{convergence}
        \STATE  $ \hat{\theta}_g =\theta_g,\hat{\theta}_c=\theta_c,\hat{\theta}_{d}=\theta_d$
        \RETURN  \bm{$\hat{\theta}_g, \hat{\theta}_c,\hat{\theta}_d$} 
        \end{algorithmic}
      \end{algorithm}

\section{Experiments}
\label{sec:Experiments}

\subsection{Datasets}
\label{ssec:subDatasets}

We systemically evaluate the proposed deep adversarial training method on the CHiME-4 corpus \cite{Vincent2016An}. Two types of data are used in this paper: `real data' that is recorded in real noisy environments (on a bus, cafe, pedestrian area, and street junction); `simulated data' that has been generated by artificially mixing clean speech data from the WSJ0 set with noisy backgrounds. The training set consists of 1,600 real and 7,138 simulated utterances in the 4 noisy environments. Each utterance consists of six channels and we randomly choose one channel signals. The development and test sets consist of 3,280 and 2,640 utterances, respectively, each containing equal quantities of real and simulated data. We choose one channel signals according to the official instructions (dt/et05\underline{\hspace{0.5em}}simu/real\underline{\hspace{0.5em}}1ch\underline{\hspace{0.5em}}track.list). In addition, the original WSJ0 training data (si\underline{\hspace{0.5em}}tr\underline{\hspace{0.5em}}s, 7,138 utterances) is used as the clean speech for GANs training.

\subsection{Setup}
\label{ssec:subSetup}

In following experiments, we take the 80-dimensional filter-banks as the input features, and each dimension of features is normalized to have zero mean and unit variance over the training set. To capture temporal information, 19 frames of context features are concatenated as the final inputs. Firstly, the splice context of noisy features $\tilde{\mathbf{x}}$ is feed into the generator G to generate its enhanced version $\hat{\mathbf{x}}$. Then the generated data used as `False' samples and the randomly chosen true clean data used as 'True' samples form the input dataset of the discriminator D. We mention that there is no need for one-to-one correspondence between the true clean and generated samples. Finally, the classifier C uses the hidden output $\mathbf{h}$ of last encoder layer of G to compute the posterior probabilities of HMM state instead of G generating enhanced data. The frame-level targets required in training come from a well-trained Gaussian Mixture Model (GMM-HMM) system. In fact, the encoding parts of G and the classifier C can be regarded as an organic whole of acoustic model.

The generator G is composed of 8 2-D strided convolutional layers. Each convolutional block consists of a convolutional layer with $3 \times 3$ filters and $2 \times 1$ strides, followed by LeakyRelu function. As mentioned, the decoder stage of G is a mirroring of the encoder with the same filter widths and the same number of filters per-layer. However, skip connections make the number of feature maps in every decoder layer to be doubled. The classifier C is a multilayer perceptron (MLP) with two hidden layers of 1024 neurons. We use rectified linear unit (Relu) \cite{Nair2010Rectified} as the activation of the hidden layer, and a softmax function of the output layer. The dropout \cite{srivastava2014dropout} with a probability of 0.3 is added across the hidden layers. And we simply use a MLP with only one hidden layer as the discriminator D. The whole system is optimized by the adversarial training algorithm (denoted as `DA-train'). We use the Adam optimizer \cite{kingma2014adam} with a minibatch of size 256 and a learning rate of 0.0002. After finishing the training, we remove the D network and only use the combination of encoder layers of G and the classifier C for testing.

The baseline system is bulit following the Kaldi chime4/s5\underline{\hspace{0.5em}}1ch recipe \cite{povey2011kaldi}(denoted as `Baseline'). The acoustic model is a DNN with 7 hidden layers. After RBM pre-training, the model is trained by minimizing the cross-entropy loss. Note that we don't perform any sequence training or LM-rescoring methods because we focus on the frame level in this paper and will investigate methods at the full-sequence level in the future.

In addition, we also compare with the same acoustic model architecture as `DA-train', but without the adversarial training. The comparison model is trained with same dataset and optimized by minimizing the cross-entropy loss (denoted as `CE-train'). In fact, `CE-train' is equivalent to the case of $\alpha = 0$ in Eq.~\ref{vgall}.

The ``CE-train" and ``DA-train" systems are implemented with PyTorch \cite{pytorch}. The WSJ 5k trigram LM is used as the language model and the Kaldi WFST decoder for decoding in all the experiments.

\subsection{Results}
\label{ssec:subResults}

\begin{table}[tb]\small
  \caption{ \small WERs (\%) of different hyper-parameter $\alpha$ on the development set}
  \label{tab:wer-alpha}
  \centering
  \begin{tabular}{p{0.7cm}<{\centering} | p{0.7cm}<{\centering} | p{0.7cm}<{\centering} | p{0.7cm}<{\centering}| p{0.7cm}<{\centering}| p{0.7cm}<{\centering}}
  \Xhline{1.5pt}
  $\alpha$ & 0.0 & 0.2 &0.4 &0.6 &0.8 \\
  \hline
  WER &18.96 &17.19 &16.32 &16.66 &16.96 \\
  \Xhline{1.5pt}
  \end{tabular}
\end{table}

\begin{table}[tb] \small
  \caption{\small WERs (\%) on the CHIME-4 corpus. `Baseline' is the model following the Kaldi chime4/s5\underline{\hspace{0.5em}}1ch. `CE-train' is the  system trained by minimizing the cross-entropy loss. `DA-train' is the proposed system optimized by deep adversarial training. Relative error rate reduction (PPER)(\%) is also shown.}
  \label{tab:Results}
  \centering
  \begin{tabular}{p{0.5cm}<{\centering} | p{0.5cm}<{\centering} | p{1.2cm}<{\centering} | p{1.2cm}<{\centering}| p{1.2cm}<{\centering}| p{1.2cm}<{\centering}}
  \Xhline{2.0pt}
  \multicolumn{2} {c|} {\thead{Data}}    & Baseline & CE-train  &  DA-train      & RERR        \\
  \hline
  \hline
  \multirowcell{3}{dev} & simu &  20.46 &  18.27    &  15.34    & 25.02 \\
  \cline{2-6}
  & real &  22.13 &  19.65    &  17.30    & 21.83 \\
  \cline{2-6}
  & avg. &  21.30 &  18.96    &  16.32    & 23.38 \\
  \Xhline{1.0pt}
  \multirowcell{3}{test} & simu &  29.94 &  27.66    &  24.75    &  17.33 \\
  \cline{2-6}
  & real &  36.27 &  34.15    &  33.83    &  6.73\\
  \cline{2-6}
  & avg. &  33.11 &  30.91    &  29.29    &  11.54\\
   \Xhline{2.0pt}
  \end{tabular}
\end{table}

Firstly, we evaluate how the hyper-parameter $\alpha$ in Eq.~\ref{vgall} affects the performance of ASR. $\alpha$ specifies the tradeoff between the adversarial loss and the category loss for the optimization objective of G. When $\alpha$ is tiny, the category loss plays a main role and the adversarial loss rarely works, which may result in the acoustic model only focusing on discriminative feature and having worse generalization. On the other hand, when $\alpha$ is huge, the model may be lack of discriminant ability with weak constraint of category loss. Thus, an appropriate $\alpha$ is very important to get better performance of ASR. We explore the different $\alpha$ while keeping the other hyper-parameters fixed in the experiments. As shown in Table~\ref{tab:wer-alpha}, the WER decreases with the increase of $\alpha$. However, when $\alpha$ reaches 0.4, the WER will increase. Therefore, we choose $\alpha= 0.4$  for the following experiments.

Table~\ref{tab:Results} reports the WER results of different recognition models on the development and test dataset. In addition, we also report the relative error rate reduction (RERR) in the last column of Table~\ref{tab:Results}. It can be seen that the proposed `DA-train' consistently and significantly outperforms the baseline (`Baseline') and comparison (`CE-train') method on both development and test set. Compared with `CE-train', the proposed `DA-train' achieves significant performance improvements (from 18.96 to 16.32 on development set and from 30.91 to 29.29 on test set). It mainly owns to the proposed deep adversarial training boost noise robustness of acoustic model. Moreover, compared with `Baseline', we achieve average RERRs of 23.38 \% and 11.54\% on the development and test set, respectively.

\begin{figure}[tb]\small
  \centering
  \includegraphics[width=0.8\linewidth, trim=25 162 40 8,clip]{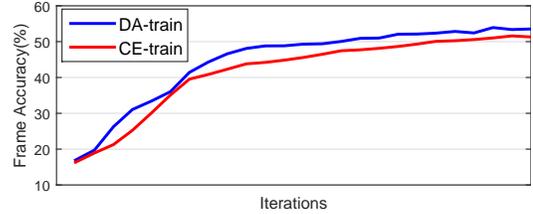}\\
  \caption{\small Frame accuracy during training for DA-train and CE-train}
  \label{fig:frame}
\end{figure}

In order to further make a comparison between deep adversarial and cross-entropy training, we analyze the frame accuracy during training. Figure~\ref{fig:frame} shows that the model trained by adversarial manner not only arrives at higher frame accuracy but also learns more quickly. These results suggest that our proposed training method works efficiently and improves ASR performance.

\section{Conclusions} \small
\label{sec:conclusions}

In this paper, we propose an adversarial training method to directly boost noise robustness of acoustic model. Specifically, a jointly compositional scheme of generative adversarial net (GAN) and neural network-based acoustic model (AM) is used in training phase. GAN is used to generate clean feature representations from noisy features by the guidance of a discriminator that tries to distinguish between true clean signals and the generated signals. The joint optimization of generator, discriminator and AM concentrates the strengths of both GAN and AM for speech recognition. Systematic
experiments on CHiME-4 show that the proposed method significantly improves the noise robustness of AM and achieves 23.38\% and 11.54\% relative improvements in WER. In the future, we will perform deep adversarial training for acoustic modeling at the full-sequence level and perform our experiments on a larger dataset.

\section{Acknowledgements}
This work was supported in part by the National Natural Science Foundation of China (No. 61573357, No. 61503382, No. 61403370, No. 61273267, No. 91120303).

{
\footnotesize
\bibliographystyle{IEEEbib}
\bibliography{strings,refs}
}
\end{document}